\documentclass[a4paper]{jpconf}
\usepackage{graphicx}
\begin{document}

\newcommand{\ee}{e^{+} e^{-}}
\newcommand{\bbar}{B\bar{B}}
\newcommand{\qqbar}{q\bar{q}}
\newcommand{\ccbar}{c\bar{c}}
\newcommand{\leplep}{\ell^{+}\ell^{-}}
\newcommand{\jp}{J/\psi}
\newcommand{\psip}{\psi '}
\newcommand{\jpsi}{J/\psi}
\newcommand{\mumu}{\mu^{+}\mu^{-}}
\newcommand{\pipi}{\pi^{+}\pi^{-}}
\newcommand{\piz}{\pi^{0}}
\newcommand{\etac}{\eta_{c}}
\newcommand{\etacp}{\eta_{c}^{\prime}}
\newcommand{\ks}{K_{S}}
\newcommand{\dsj}{D_{sJ}}
\newcommand{\kstr}{K^{*0}}
\newcommand{\B}{B^{0}}
\newcommand{\Bp}{B^{+}}
\newcommand{\ecp}{\eta_{c}(2S)}
\newcommand{\kpi}{K^-\pi^{+}}
\newcommand{\Mbc}{M_{\rm bc}}
\newcommand{\DE}{\Delta E}
\newcommand{\fb}{fb^{-1}}
\newcommand{\ra}{\rightarrow}
\newcommand{\rt}{\rightarrow}
\newcommand{\etalus}{\em et al.}

\title{New particles from Belle}

\author{Stephen L. Olsen}

\address{Department of Physics \& Astronomy, University of Hawai'i at Manoa, Honolulu, HI 96822, 
USA}

\ead{solsen@phys.hawaii.edu}

\begin{abstract}
I report recent results on hidden charm spectroscopy from Belle.
These include: 
observation of a near-threshold enhancement 
in the  $\omega\jp$ invariant mass distribution for exclusive 
$B\rt K \omega\jp$ decays; 
evidence for the decay $X(3872)\rt\pipi\piz\jp$,
where the $\pipi\piz$ invariant mass distribution has a
strong peak between 750~MeV and the kinematic limit of 775~MeV,
suggesting that the process is dominated by the sub-threshold
decay $X\rt\omega\jp$;  
and the observation of a peak
near 3940~MeV in the $\jp$ recoil mass spectrum for the
inclusive continuum process $\ee\rt\jp X$.
The results are based on a study of a 287~fb$^{-1}$ sample
$\ee$ annihilation data collected at center-of-mass energies around the
$\Upsilon(4S)$ in the Belle detector at 
the KEKB collider.  
 
\end{abstract}.

\section{Introduction}
The recent surge in activity in hadron spectroscopy and, I suppose, the
main motivation for the formation of the Topical Group on Hadron Physics,
is the result of renewed interest in a rather old question:  
{\em are there hadronic states  with
a more complex structure than the simple $\qqbar$ mesons and $qqq$ baryons
of the original quark model}?  This revival of interest has been driven by 
experimental reports
of pentaquarks~\cite{pentaquark}, the narrow $\dsj$ 
states~\cite{D_sJ,pavel_dsj}, and the $X(3872)$~\cite{skchoi_x3872}.

In spite of considerable theoretical and experimental effort, the existence
of non-$\qqbar$ mesons and/or non-$qqq$ baryons remains an open question.  
While the identification of a strangeness=+1 (or charm=-1)
baryon would be definitive evidence for a non-$qqq$ baryon,
the experimental situation regarding the existence of such  
states remains unsettled (and a major topic of discussion at
this meeting~\cite{dzierba}). On the other hand, while the
$\dsj$ and and $X(3872)$ are experimentally well established,
the theoretical interpretation is not so clear.
The $D_{sJ}$ states could be standard $P$-wave $c\bar{s}$ states 
and their narrowness is only surprising because the relativistic 
potential model calculations that predicted them to be heavier 
(and above $DK$ threshold) are wrong~\cite{eichten}.  Some theorists, 
including our opening speaker~\cite{quigg}, remain hopeful that 
a $\ccbar$ charmonium assignment can be found for the $X(3872)$.

To sort this all out, I think that the so-called hidden charm 
mesons can and will play a decisive role for reasons that include:

\begin{itemize}

\item the theory for these systems is well founded 
      (and recently blessed by this year's Nobel Prize Committee) 
      and has fewest ambiguities;

\item the experimental signatures tend to be clean;

\item $\ccbar$ meson states below open-charm 
      threshold are narrow and do not overlap; and

\item lots of non-$\ccbar$-type mesons have been conjectured,
      including $D\bar{D^*}$ molecules~\cite{molecule} and 
      $\ccbar$-$gluon$ hybrids~\cite{mandula}.

\end{itemize}

Although the Belle detector~\cite{Belle} is specialized to studies of
CP violation in $B$ meson decays, it has proven to be a useful
device for studying particles containing $\ccbar$ pairs.
Belle detects $\ccbar$ systems
produced via weak decays of $b$ quarks ---the $b\rt \ccbar s$
process is a dominant $b$-quark decay mode--- and the continuum 
production process $\ee\rt \ccbar\ccbar$, which has been
found to be surprisingly large.   The KEKB asymmetric
energy $\ee$ collider~\cite{KEKB} operates at a center-of-mass (cms)
energy corresponding to the $\Upsilon(4S)$ resonance and
routinely delivers luminosities that are in excess of 
$10^{34}$cm$^{-2}$s$^{-1}$, thereby providing  Belle with a huge 
data sample that contains  about 300 million $\bbar$ meson 
pair events and over one billion $\ee\rt\qqbar$ continuum
annihilation events.

Belle results in the hidden charm meson
sector include first observations of:

\begin{itemize}

\item the $\etacp$ via the sequence
      $B\rt K\etacp$, $\etacp\rt \ks K\pi$~\cite{skchoi_etacp};

\item anomalously large cross sections for
      the exclusive process $\ee\rt\jp\etac$ and the inclusive 
      process $\ee\rt\jp (\ccbar)$~\cite{pakhlov_sigmaetac}; 

\item the $X(3872)$ meson~\cite{skchoi_x3872};

\item a near-threshold $\omega\jp$ mass enhancement
      in exclusive $B\rt K\omega\jp$ decays~\cite{kabe_omegajpsi}; and

\item a peak at 3940~MeV in the $\jp$ recoil mass
      spectrum in the inclusive $\ee\rt\jp X$ process~\cite{pakhlov_jpetac}.

\end{itemize}

In this talk I will discuss the last two items as well as 
recent results on properties of the $X(3872)$.  I will
not have time to cover any of the many other Belle results 
on hadron spectroscopy, such as our many interesting results 
on charmed baryon spectroscopy~\cite{lambda_c}, $D^{**}$~\cite{kuzmin} and 
$\dsj$ mesons~\cite{pavel_dsj} and two-photon physics~\cite{twophot}.  In 
addition, I will not have
time to report on Belle's lack of observation of
pentaquarks~\cite{mizuk} or the $D_{sJ}(2632)$~\cite{kabe_2632}. 
All unpublished numbers reported here are preliminary.

\section{A near-threshold $\omega\jp$ 
mass enhancement in $B \rt K \omega\jp$ decays}

At the $\Upsilon(4S)$, $\bbar$ meson pairs are
produced with no accompanying particles. As a result, each
$B$ meson has a total cms 
energy that is equal to $E_{\rm beam}$, the cms beam energy.  
We identify $B$ mesons using the beam-constrained 
$B$-meson mass 
$\Mbc=\sqrt{E_{\rm beam}^2 - p_B^2}$ and the energy difference
$\DE = E_{\rm beam} - E_B$, where $p_B$ is the vector sum of the 
cms momenta of the $B$ meson decay products and $E_B$ is their
cms energy sum.  For the final states discussed here,
the experimental resolutions for $\Mbc$ and $\DE$ are
approximately 3~MeV and 13~MeV, respectively.

\begin{figure}[h]
\begin{minipage}{18pc}
\includegraphics[width=18pc]{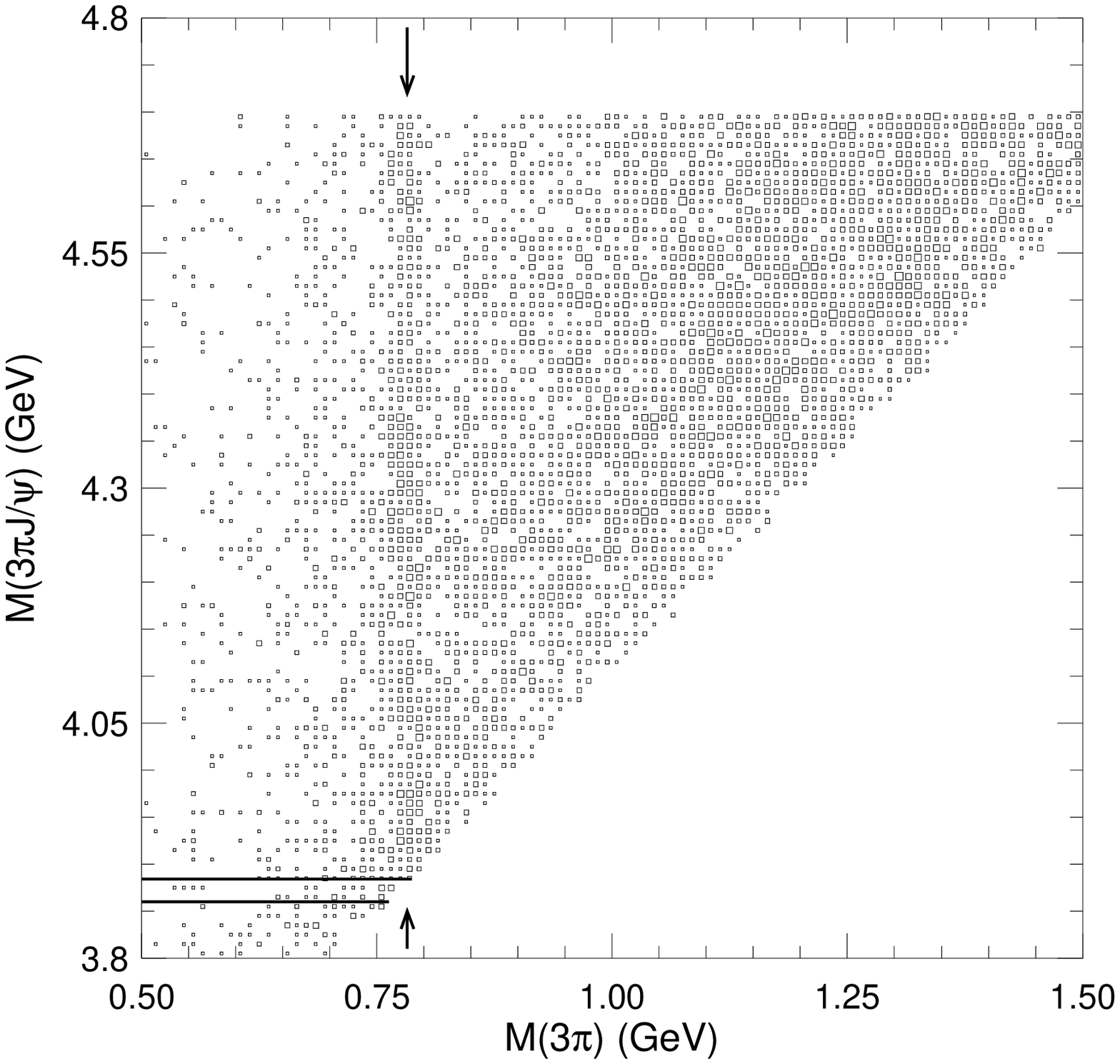}
\caption{\label{fig:3pill_vs_3pi}A scatterplot of $M(\pipi\pi^0\leplep)$ (vertical)
$versus$ $M(\pipi\pi^0)$ for events
in the $\DE$-$\Mbc$ signal region.  The vertical band
indicated by the arrows corresponds to $\omega\rt\pipi\pi^0$
decays.}
\end{minipage}\hspace{2pc}%
\begin{minipage}{18pc}
\includegraphics[width=18pc]{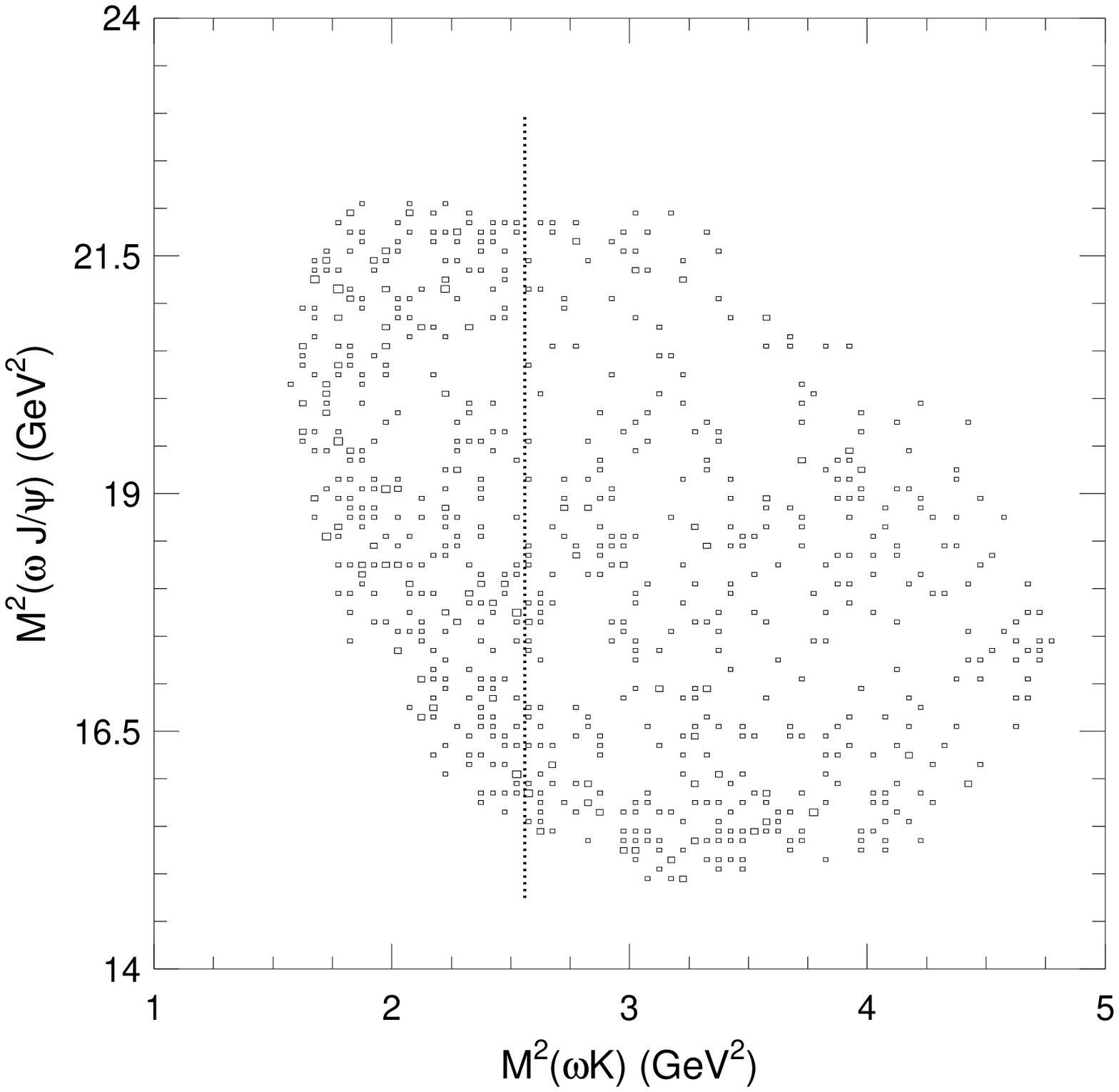}
\caption{\label{fig:dalitz}the Dalitz-plot distribution for 
$B\rt K\omega\jp$ candidate events.
}
\end{minipage} 
\end{figure}

We select $B\rt K\pipi\pi^0 \jp$ candidate events
($\jp\rt\leplep)$ track combinations with
$\Mbc$ and $\DE$ values that are within $2.5\sigma$ 
of their nominal values.  
Figure~\ref{fig:3pill_vs_3pi}
shows a scatterplot of $M(\pipi\pi^0\jp)$ (vertical)
$versus$ $M(\pipi\pi^0)$ for selected events in the
$\DE$-$\Mbc$ signal region.
Here a distinct vertical band corresponding to $\omega\rt\pipi\pi^0$ decays is
evident near $M(\pipi\pi^0) = 0.782$~GeV.

We identify three-pion combinations with
$M(\pipi\pi^0)$ within 25~MeV of $m_{\omega}$ as
$\omega$ candidates and form the Dalitz plot of
$M^2(\omega\jp)$ (vertical) $versus$ $M^2(\omega K)$
(horizontal) shown in Fig.~\ref{fig:dalitz}.  The
clustering of events near the left side of the plot
corresponds to $B\rt K_X\jp$; $K_X \rt K\omega$ events,  
where $K_X$ denotes strange meson resonances such as  
$K_1(1270)$, $K_1(1400)$, and $K^*_2(1430)$ that are 
known to decay to $K\omega$.  There is also a 
clustering of events with low $\omega\jp$ invariant 
masses  near the bottom of the Dalitz plot.  To study 
these, we suppress $K_X\rt K\omega$ events by only looking
at events in the region $M(K\omega)>1.6$~GeV, to the 
right of the dashed line in Fig.~\ref{fig:dalitz}.

\begin{figure}[h]
\begin{minipage}{18pc}
\includegraphics[width=18pc]{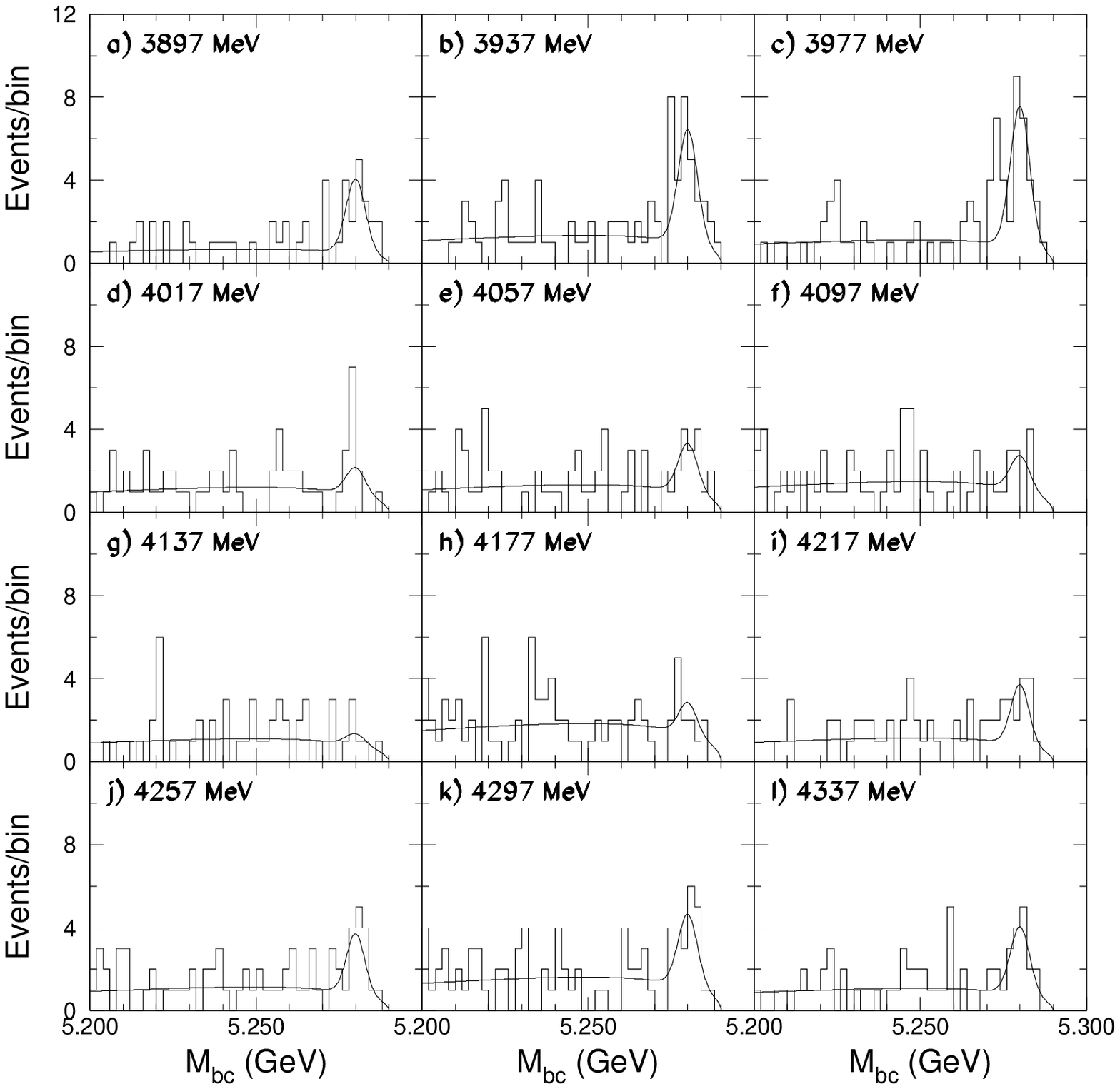}
\caption{\label{fig:mb_12box}
$\Mbc$ distributions for events in the
$\DE$ signal region for $40$~MeV-wide bins
in $M(\omega\jp)$. 
}
\end{minipage}\hspace{2pc}%
\begin{minipage}{18pc}
\includegraphics[width=18pc]{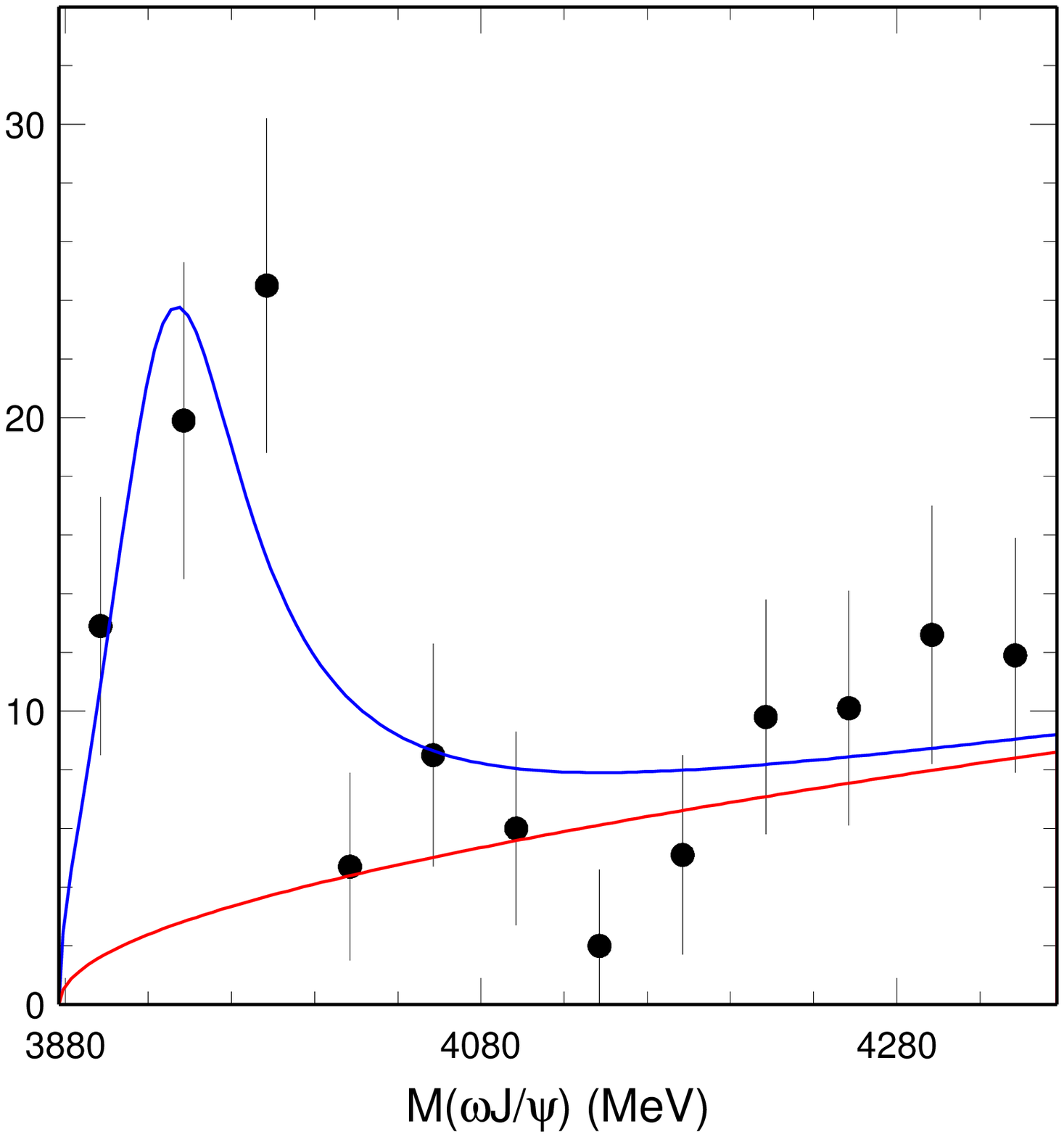}
\caption{\label{fig:slice_fits}
$B\rt K\omega\jp$ signal yields
$vs$ $M(\omega\jp).$  The curve in {\bf (a)}
indicates the result of a fit that 
uses an
$S$-wave Breit-Wigner resonance term and
a phase-space-like threshold function for
the background.
}
\end{minipage} 
\end{figure}

The  $\Mbc$ and $\DE$ distributions of the selected
events indicate that about half of the entries in the
$M(\omega K)>1.6$~GeV Dalitz plot region are
due to background.  To perform a background
subtraction and determine the level of
$B\rt K\omega\jp$ signal events, we separate
the data into $40$~MeV-wide bins of $M(\omega\jp)$
and measure the $B$ meson signal levels in the $\Mbc$
and $\DE$ distributions.  The histograms in Fig.~\ref{fig:mb_12box}
show the $\Mbc$ distributions for the twelve lowest
$M(\omega\jp)$ mass bins, where strong peaks at 
$\Mbc=m_B$ are evident at low $\omega\jp$
masses,  especially for the mass regions
covered by Figs.~\ref{fig:mb_12box}(b) and (c). 
The corresponding $\DE$ distributions
(not shown) show similar structure.
We establish the $B\rt K\omega\jp$ signal level
for each $M(\omega\jp)$ mass bin by performing
binned fits simultaneously to the $\Mbc$ and $\DE$ 
distributions with Gaussian functions for the signal
and smooth background functions.  The smooth
curves in Fig.~\ref{fig:mb_12box} indicate the
fitted $\Mbc$ curves for each $\omega\jp$ mass bin.

The bin-by-bin signal yields
are plotted $vs$ $M(\omega\jp)$
in Fig.~\ref{fig:slice_fits}.
An enhancement is evident around 
$M(\omega\jp)=3940$~MeV.
The curve in  Fig.~\ref{fig:slice_fits}
is the result of a fit with a
$S$-wave Breit Wigner function
threshold function of the form $f(M) = A_0 q^*(M)$,
where $q^*(M)$ is the momentum of the daughter
particles in the $\omega \jp$ restframe.
This functional form accurately reproduces
the threshold behavior of Monte Carlo simulated  
$B\rt K\omega\jp$ events that are generated 
uniformly distributed over phase-space.

The fit gives a Breit-Wigner signal yield of 
$58\pm 11$ events with a peak
peak mass and total width of
\begin{eqnarray*}
M       &=&    3943 \pm 11 {\rm (stat)} \pm 13 {\rm (syst)}~{\rm MeV}\\
\Gamma  &=&     87  \pm 22 {\rm (stat)} \pm 26 {\rm (syst)}~{\rm MeV},
\end{eqnarray*}
where the systematic errors are determined from variations 
in the values when different bin sizes, fitting shapes and
selection criteria are used.
The event yield translates into 
a product branching fraction (here we denote the enhancement as $Y(3940)$):
\[
{\cal B}(B\rt K Y(3940)){\cal B}(Y(3940)\rt\omega\jp) = 
(7.1 \pm 1.3 {\rm (stat)} \pm 3.1 {\rm (syst)})\times 10^{-5},
\]
The statistical significance
of the signal, determined from
$\sqrt{-2\ln({\mathcal L}_0/{\mathcal L}_{\rm max})}$,
where ${\mathcal L}_{\rm max}$ and ${\mathcal L}_0$ are the likelihood
values for the best-fit and for zero-signal-yield, respectively, 
is $8.1\sigma$.

A $\ccbar$ charmonium meson a mass of 3943~MeV would dominantly
decay to  $D\bar{D}$ and/or $D\bar{D}^*$;  hadronic
charmonium transitions should have minuscule branching
fractions.  On the other hand, decays of $c\bar{c}$-$gluon$  
hybrid charmonium  to $D^{(*)}\bar{D^{(*)}}$ meson 
pairs are forbidden or suppressed, and the relevant ``open 
charm'' threshold is  
$m_D + m_{D^{**}} \simeq 4285$~MeV~\cite{isgur,page},  
where $D^{**}$ refers to the $J^P = (0,1,2)^+$ charmed mesons.
Thus, a hybrid state with a mass equal to that of the peak 
we observe would have large branching fractions for decays
to $\jp$ or $\psip$ plus light hadrons~\cite{close}.  
Moreover, lattice QCD calculations have indicated 
that partial widths for such decays can be comparable 
to the width that we measure~\cite{michael}.  However,
these calculations predict masses for these states
that are between 4300 and 4500~MeV~\cite{banner}, 
substantially higher than our measured value.

\section{The $X(3872)$ with 253 fb$^{-1}$}

\begin{figure}[h]
\begin{minipage}{18pc}
\includegraphics[width=18pc]{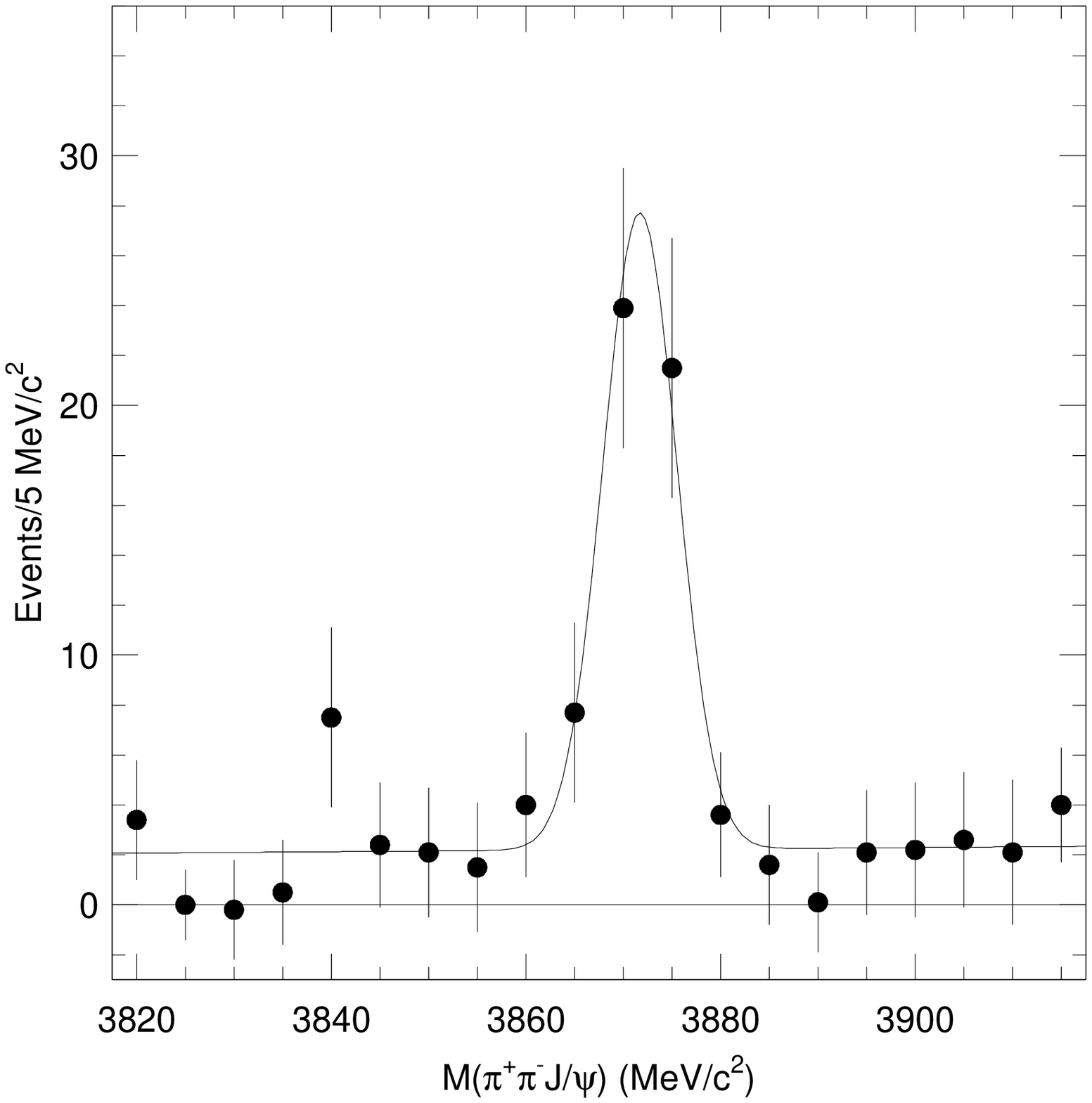}
\caption{\label{fig:Mpipijpsi}
The $X(3872)\rt\pipi\jp$ signal from the 253~fb$^{-1}$
data sample.}
\end{minipage}\hspace{2pc}%
\begin{minipage}{18pc}
\includegraphics[width=18pc]{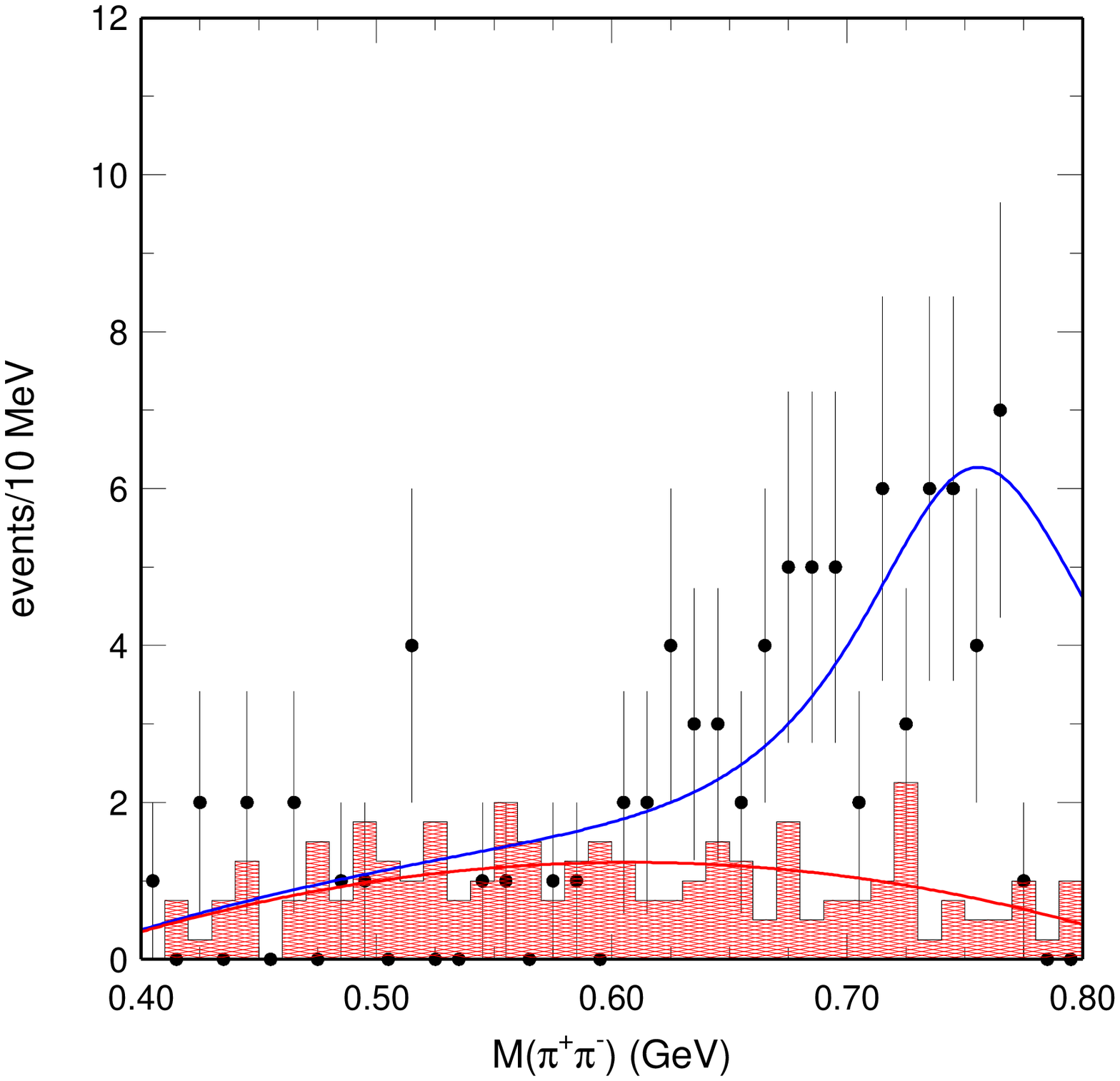}
\caption{\label{fig:Mpipi}$M(\pipi)$ for events in the 
$X(3872)$ signal peak.  The shaded histogram is the
sideband-determined background; the curve is the result 
of a fit with a $\rho\rt\pipi$ lineshape.}
\end{minipage} 
\end{figure}

The $X(3872)$ was  discovered by Belle as a 
narrow $\pipi\jp$ mass peak in exclusive 
$B^-\rt K^-\pipi\jp$ decays~\cite{skchoi_x3872,conj}.  
Figure~\ref{fig:Mpipijpsi} shows the $X(3872)$ signal
from a 253~fb$^{-1}$ data sample containing 275 million $\bbar$ pairs.
The observed mass and the narrow width are not compatible
with expectations for any of the as-yet unobserved 
charmonium states~\cite{solsen_krakow}.  Moreover, the
$\pipi$ invariant mass distribution,  shown in
Fig.~\ref{fig:Mpipi}, peaks
near the upper kinematic limit of $M(\pipi)=775$~MeV, and
has a shape that is consistent with $\rho\rt\pipi$ decays.  
Charmonium decays to 
$\rho\jp$ final states violate isospin and are 
expected to be  suppressed.  The $X(3872)$ and its above-listed 
properties were confirmed by the BaBar~\cite{X_BaBar}, 
CDF~\cite{X_CDF} and D0~\cite{X_D0} experiments.

The $X(3872)$ mass ($3871.9\pm 0.5$~MeV~\cite{X_mass}) is within
errors of the  $D^0\bar{D^{0*}}$ threshold ($3871.3\pm 
1.0$~MeV~\cite{PDG}); 
the difference is $0.6\pm 1.1$~MeV. This
has led to speculation that the $X$ might be a $D^0\bar{D^{0*}}$ 
bound state~\cite{tornqvist,swanson_1,molecule}. According to 
ref.~\cite{tornqvist}, the preferred quantum numbers for such a 
bound state would be either $J^{PC} = 0^{-+}$ or $1^{++}$.  
The decay of an $C=+1$ state to $\pipi\jp$ would proceed via 
an $I=1$ $\rho^0\jp$ intermediate state and produce the
$\pipi$ mass spectrum like that we see.  In this
meson-meson bound state interpretation, the
close proximity of the $X$ mass to $D^0\bar{D^{0*}}$ threshold
compared to the $D^+ D^{-*}$-$D^0\bar{D^{0*}}$ mass splitting
of 8.1~MeV produces a strong isospin violation.

Swanson made a dynamical model for the $X(3872)$ as a 
$D^0\bar{D^{0*}}$ hadronic resonance~\cite{swanson_1}.  
In this model, $J^{PC}=1^{++}$ is strongly favored and
the wave function has, in addition to $D^0\bar{D^{0*}}$,
an appreciable admixture of $\omega\jp$ plus
a small $\rho\jp$ component.  The latter produces
the $\pipi\jp$ decays that have been observed; the
former gives rise to $\pipi\piz\jp$ decays via a
virtual $\omega$ that are enhanced because of
the large $\omega\jp$ component to the wavefunction.
Swanson's model predicts that
$X(3872)\rt \pipi\piz\jp$ decays should
occur at about half the rate for $\pipi\jp$ and
with a $\pipi\piz$ invariant mass spectrum that
peaks near the upper kinematic boundary of 775~MeV
(7.5~MeV below the $\omega$ peak).

$X(3872)\rt\pipi\piz\jp$ decays would populate the
horizontal band indicated by the horizontal lines
in the scatterplot of Fig.~\ref{fig:3pill_vs_3pi}.
This corresponds to the $\pm 3\sigma$ band
$|M(\pipi\piz\jp)-m_{X(3872)}|<16.5$~MeV.

\begin{figure}[h]
\begin{minipage}{18pc}
\includegraphics[width=18pc]{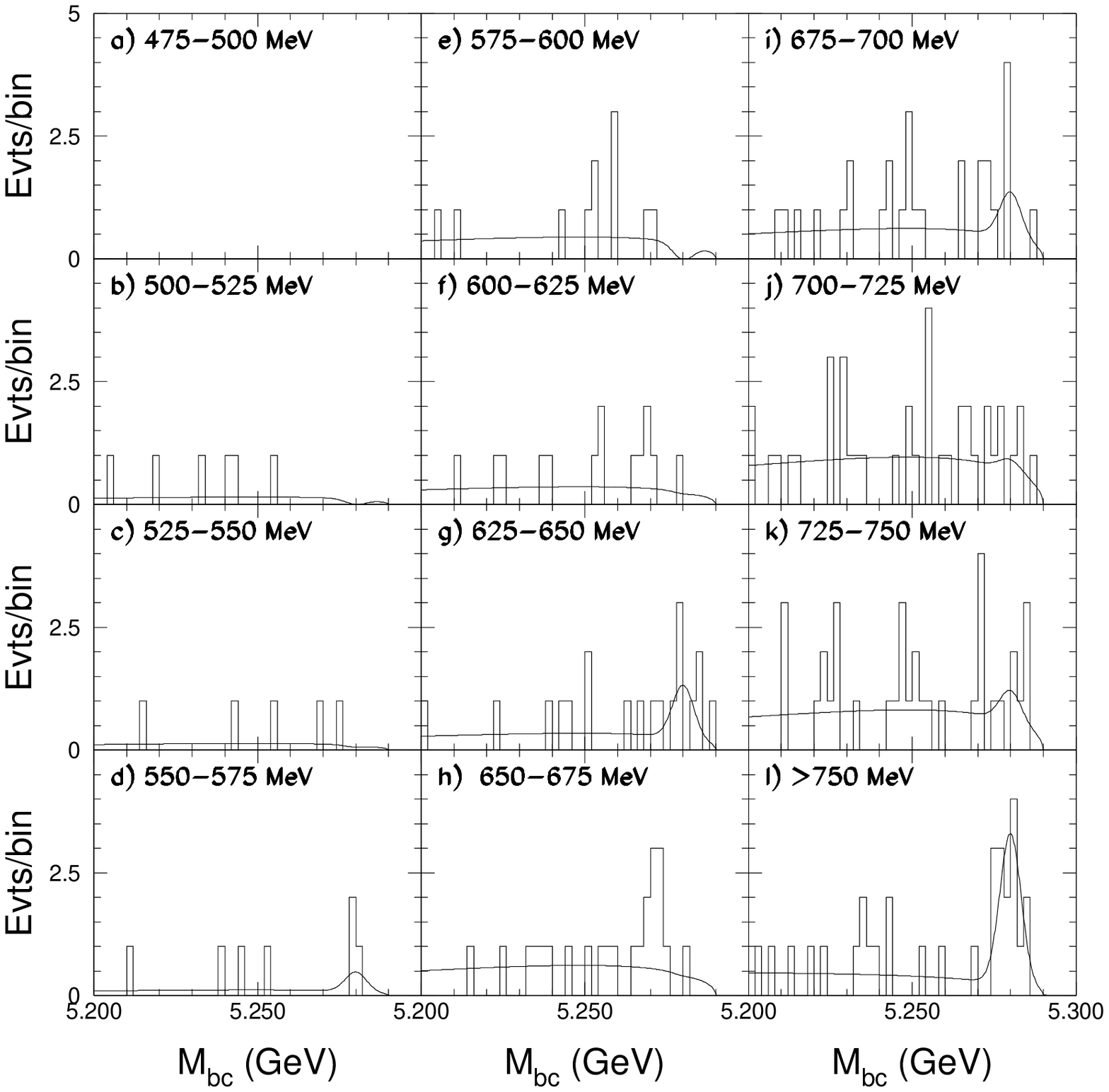}
\caption{\label{fig:x3pi_mb_12box}
$\Mbc$ distributions for 25 MeV-wide $\pipi\piz$
invariant mass bins.
}
\end{minipage}\hspace{2pc}%
\begin{minipage}{18pc}
\includegraphics[width=18pc]{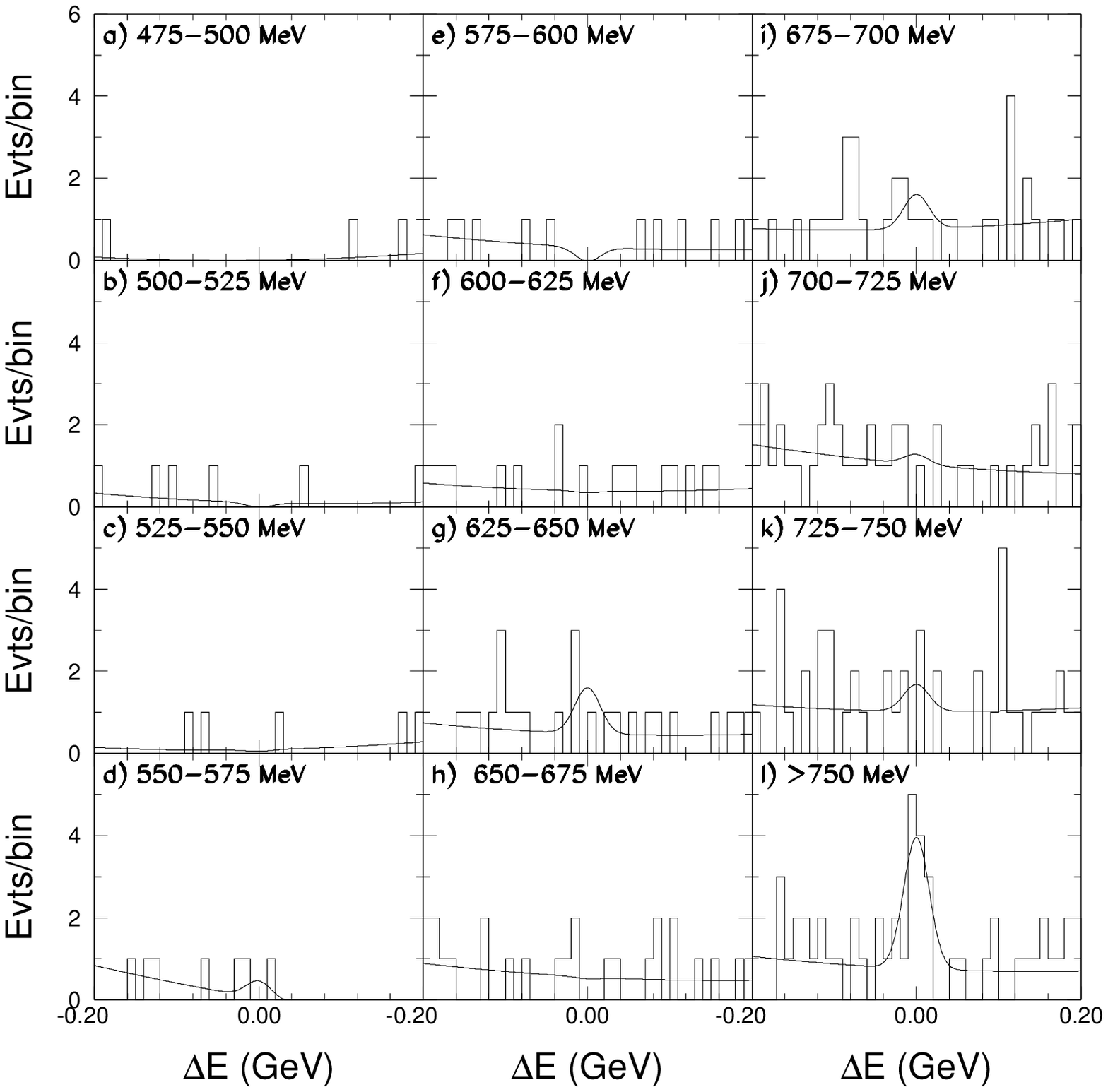}
\caption{\label{fig:x3pi_de_12box}
$\DE$ distributions for 25 MeV-wide $\pipi\piz$
invariant mass bins.
}
\end{minipage} 
\end{figure}

Figure~\ref{fig:x3pi_mb_12box} shows the $\Mbc$ 
distributions
for events that are in the $\DE$ and $X\rt\pipi\piz\jp$
signal regions for 25~MeV-wide  $\pipi\piz$
invariant mass bins; Fig.~\ref{fig:x3pi_de_12box}
shows the 
corresponding $\DE$ distributions for events
in the $\Mbc$ and $X$ signal regions.
There are distinct $B$ meson
signals in both the $\Mbc$ and $\DE$ distributions
for the $M(\pipi\piz)>750$~MeV bin and
no evident signals for any of the other $3\pi$ mass bins. 
The curves in the figures are the results
of binned likelihood fits that are applied simultaneously
to the $\Mbc$ and $\DE$ distributions.

\begin{figure}[h]
\begin{minipage}{18pc}
\includegraphics[width=18pc]{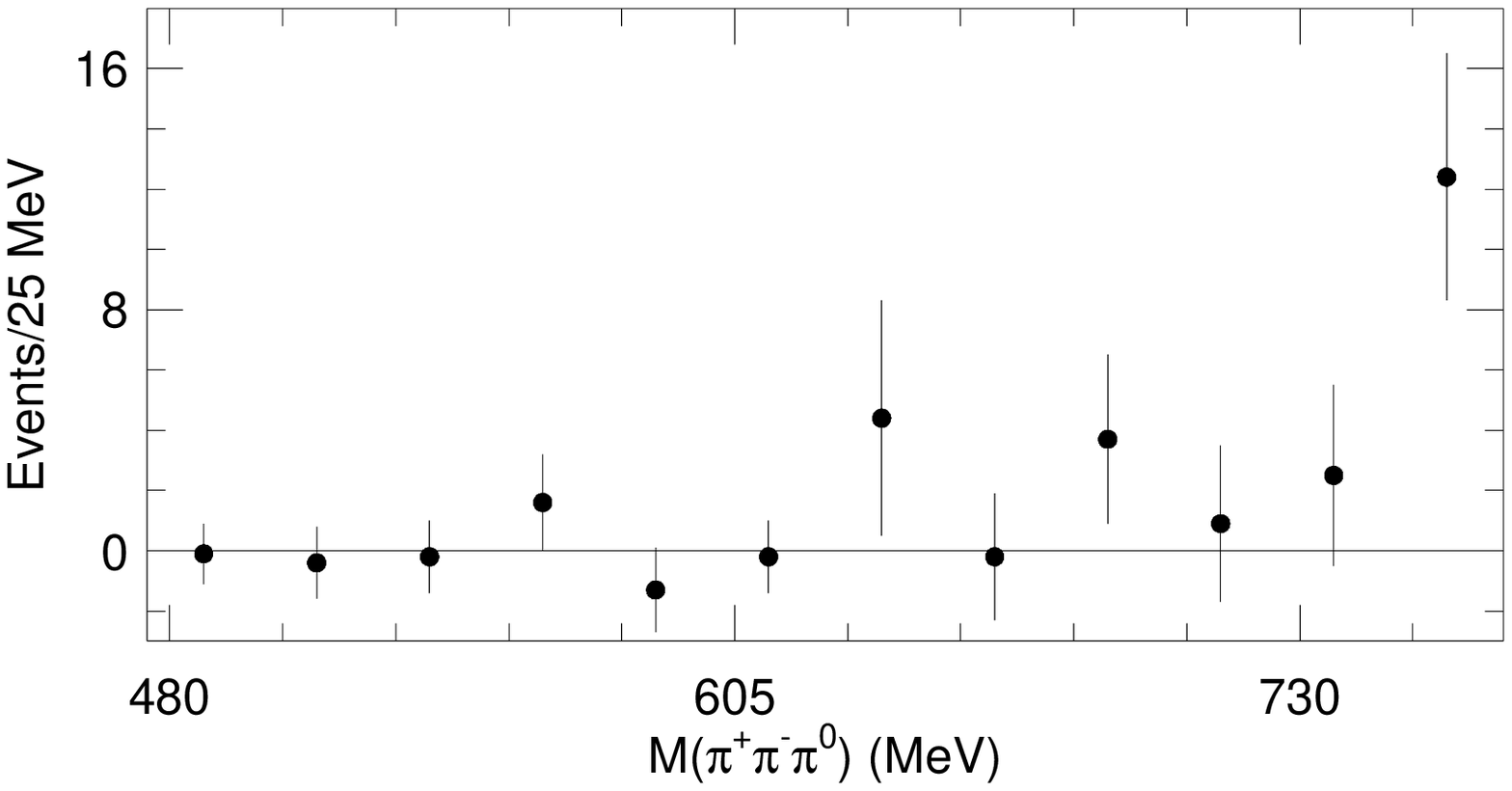}
\caption{\label{fig:m3pi}
The $B$-meson signal
yields from the fits to the
$\Mbc$-$\DE$ signals $vs$
$3\pi$ invariant mass.
}
\end{minipage}\hspace{2pc}%
\begin{minipage}{18pc}
\includegraphics[width=18pc]{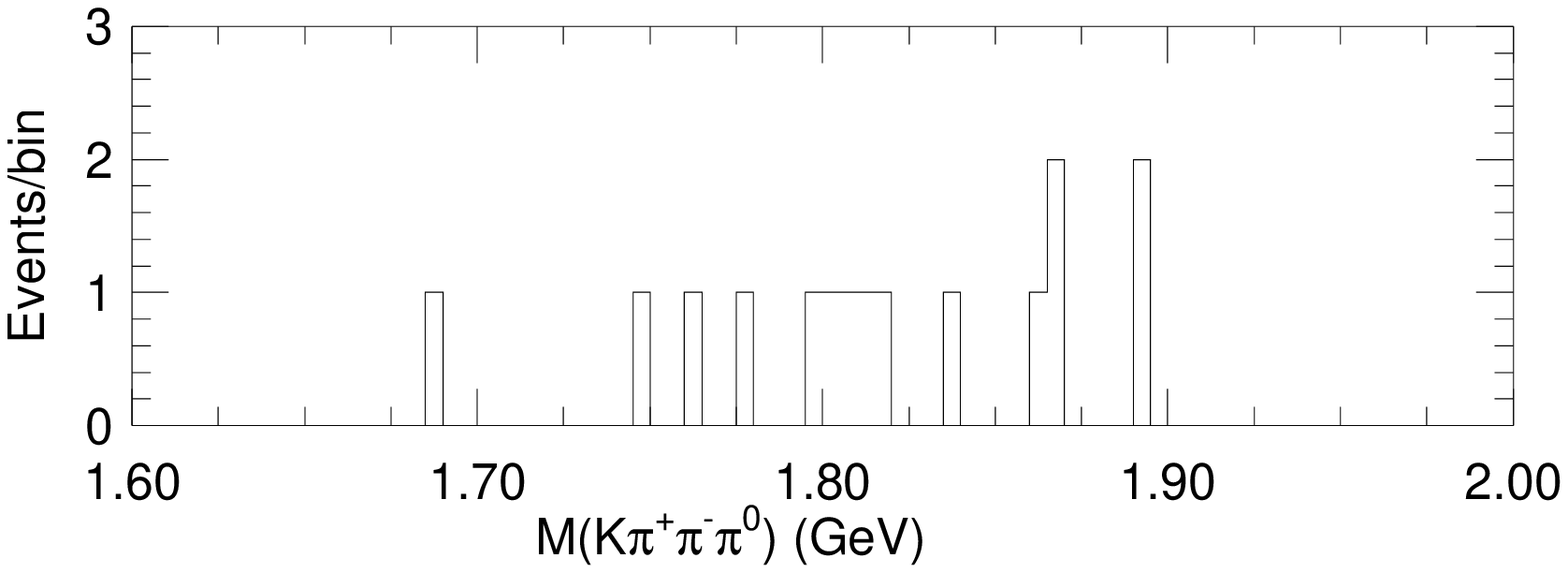}
\caption{\label{fig:mk3pi}
The $M(K\pipi\piz)$
distribution for events in the
$M_X$-$M(3\pi)$ signal region.
}
\end{minipage} 
\end{figure}

Figure~\ref{fig:m3pi} shows the fitted $B$-meson 
signal yields $vs$ $M(\pipi\pi^0)$.
All of the fitted yields are consistent
with zero except for the $M(\pipi\piz)>750$~MeV
bin, where the fit gives $12.4\pm 4.1$ events.
The statistical significance
of the signal in this one bin, determined from
$\sqrt{-2\ln({\mathcal L}_0/{\mathcal L}_{\rm max})}$,
where ${\mathcal L}_{\rm max}$ and ${\mathcal L}_0$ are the likelihood
values for the best-fit and for zero-signal-yield, respectively, 
is $6.6\sigma$.   

Figure~\ref{fig:mk3pi} shows the $M(K\pipi\piz)$ 
distribution for for the $X\rt\pipi\piz\jp$ signal events.
The distribution is spread across the limited allowed
kinematic region and there is no evident
structure that might be producing the high
mass peak in Fig.~\ref{fig:m3pi} by some sort
of a kinematic reflection.

A possible background to the observed signal would
be feed-across from the near-threshold $\omega\jp$ 
enhancement in $B\rt K\omega\jp$ decays described above.
Since the $\omega\rt\pipi\piz$
resonance peak is at $m_{\omega}=782.5$~MeV, which is 7.5~MeV above
the maximum possible $3\pi$ invariant mass value
for $X\rt\pipi\piz\jp$ decays, there is no overlap
between the centroids of the $\omega\jp$
and $X\rt\pipi\piz\jp$ signal bands in Fig.~\ref{fig:3pill_vs_3pi}.
However, there is some overlap in the tails of the kinematically 
allowed regions for the two processes that might result in some
events from one signal feeding into the other.

We determine the level of signal cross-talk to be
$0.75\pm0.14$~events from the integral 
of the fitted function in Fig.~\ref{fig:slice_fits}
over the region of overlap with the $X(3872)$ signal band.
As an independent check, we refitted for the  $X(3872)\rt\pipi\piz\jp$
signal yield with a tighter restriction on $M(\pipi\piz\jp)$, namely
$m_X - 3\sigma < M(\pipi\piz\jp) < m_X + 1\sigma$,  that has
{\em no overlap} with the $\omega$ band.
The $X\rt\pipi\piz\jp$ signal yield in the truncated region is 
$10.6\pm 3.6$ events.  For a Gaussian signal distribution with
no feed-across background, we expect the truncation of the signal region
to reduce the signal by 2.1~events (16\%); the observed reduction of 
1.8~events is consistent with a feed-across level that
is less than one event.

Another possible source of background to the $X(3872)\rt\pipi\piz\jp$
signal are non-resonant $B^-\rt K^-\pipi\piz\jp$ 
decays.  To determine the level of these,  we looked for $B$-meson 
signals in the $\Mbc$-$\DE$ distributions for events in 
$M(\pipi\piz\jp)$ sidebands
above and below the $X(3872)$ mass region. 
There is no evidence for significant signal yields in
the $\Mbc$-$\DE$ distributions of either sideband.  
Fits gives an estimate of the non-resonant background 
in the $X\rt\pipi\piz\jp$ signal bin of $1.3\pm 1.0$ events.

To determine the branching fraction, 
we attribute all of the signal events 
with $M(\pipi\piz)>750$~MeV to $X\rt\pipi\piz\jp$ decay. We
compute the ratio of $\pipi\piz\jp$ and $\pipi\jp$ branching
fractions by comparing this to the number of  $X\rt\pipi\jp$
in the same data sample, corrected by MC-determined 
relative detection efficiencies.  
The ratio of branching fractions is 
\begin{equation}
\frac{{\cal B}(X\rt\pipi\piz\jp)}{{\cal B}(X\rt\pipi\jp)}=
\frac{N_{ev}(\pipi\piz\jp)\varepsilon_{\pipi\jp}}
{N_{ev}(\pipi\jp)\varepsilon_{\pipi\piz\jp}}
= 1.1\pm0.4{\rm (stat)}\pm 0.3 {\rm (syst)},
\end{equation}
where the systematic error reflects the uncertainty in the 
relative acceptance, the level of possible feed-across and
nonresonant background, and possible event loss
due to the $M(3\pi)>750$~MeV requirement, all added in quadrature.  
If we allow for cross-talk and non-resonant contributions
at their maximum ($+1\sigma$) values, the statistical
significance of the $X(3872)\rt\pipi\piz\jp$ signal
is reduced to $\simeq 4\sigma.$

\section{A new charmonium state in inclusive $\ee\rt\jp X$
annihilations.}

\begin{figure}[h]
\begin{minipage}{18pc}
\includegraphics[width=18pc]{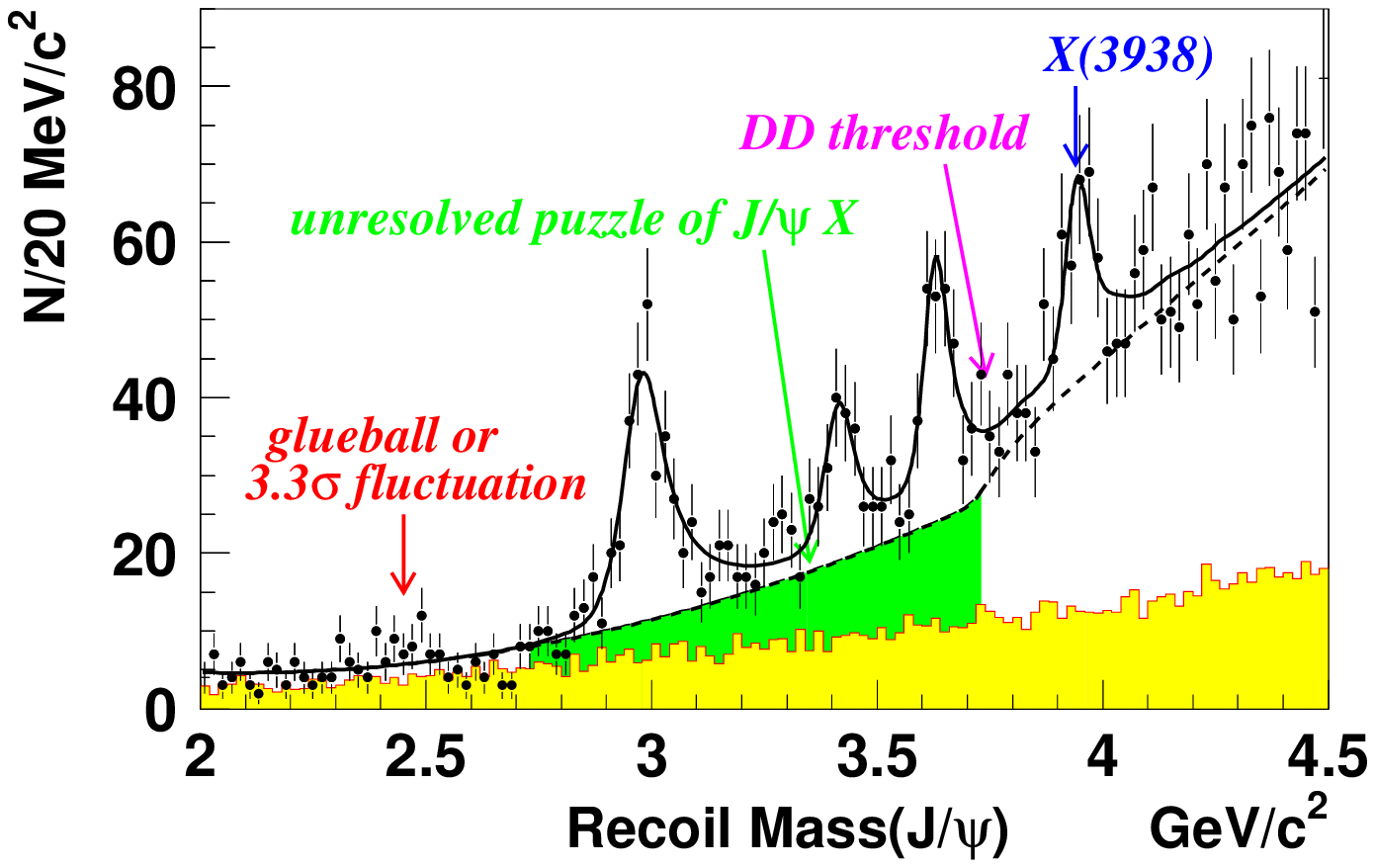}
\caption{\label{fig:jpsi_recoil}
The distribution of masses recoiling from the $\jp$
in $\ee\rt\jp X$ annihilations.
}
\end{minipage}\hspace{2pc}%
\begin{minipage}{18pc}
\includegraphics[width=18pc]{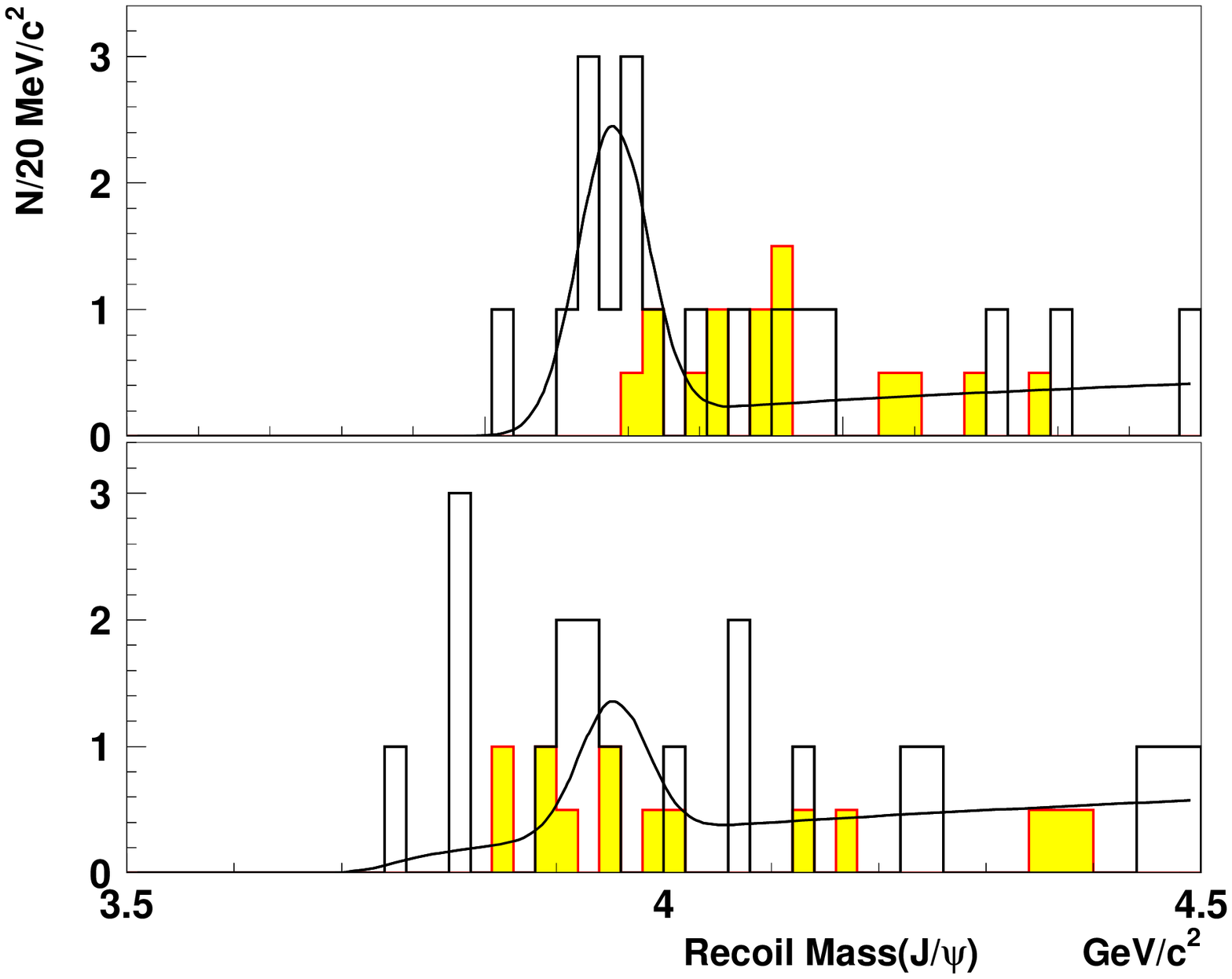}
\caption{\label{fig:Mrecoil_psi}
The $D\bar{D^*}$ (top) and
$D\bar{D}$ (bottom) invariant mass distributions
for $\ee\rt\jp D\bar{D^{(*)}}$.
}
\end{minipage} 
\end{figure}

Some of the biggest surprises from Belle have nothing to
do with $B$-meson physics at all and have come, instead, from the 
inclusive $\ee\rt\jp X$ annihilation process.  This is
demonstrated in Fig.~\ref{fig:jpsi_recoil}, which shows the 
distribution of masses for systems with more than
two charged tracks that recoil against $\jp$ mesons 
produced in the $\ee$ continuum at or near the $\Upsilon(4S)$
resonance.  In this figure, which is based on a 280~fb$^{-1}$ 
 data sample, the histogram indicates the background
level derived from the $\jp\rt\leplep$ mass sidebands.

The prominent peak at $M_{\rm recoil} \simeq 2.98$~GeV 
in Fig.~\ref{fig:jpsi_recoil} corresponds to the
$\etac$. From the yield of events we determine
a cross-section branching-fraction product~\cite{pakhlov_jpetac}
\[
\sigma_{\rm Born}(\ee\rt\jp\etac){\cal{B}}(\etac\rt >2tracks) = 25.6 \pm 4.4~{\rm fb.}
\]
This is more than an order of magnitude higher than non-relativistic
QCD (NRQCD) calculations of $\sim2$~fb$^{-1}$\cite{braaten}.
There is no evident signal for any recoils with mass below
$M_{\etac}$, which is also the $\ccbar$ mass threshold. 
Also contrary to NRQCD expectations, the four-charmed-quark
process $\ee\rt\ccbar\ccbar$ dominates inclusive $\jp$ production.
From the total number of charmonium states and charmed particles
found in the recoil system, we determine the  cross section ratio~\cite{belle_conf_0331}
\[
\frac{\sigma(\ee\rt\jp (\ccbar)}{\sigma(\ee\rt\jp X)} = 0.82\pm 0.21;
\]
NRQCD predicts this ratio to be $\sim 0.1$~\cite{cho}.

The second and third prominent  peaks in Fig.~\ref{fig:jpsi_recoil} 
are at the masses of the $\chi_{c0}$ and $\etacp$, respectively.
The fourth peak is well fitted by a Gaussian function
with a peak mass of  $3940\pm 12$~MeV and a signal significance
of $4.5\sigma$.  The width of this state is consistent with
experimental $M_{\rm recoil}$ resolution.  Since this is  
rather poor, we can only derive an upper limit on the 
total width of $\Gamma < 96$~MeV~(90\%~CL).

We investigated this peak further by studying events
where a $D$ meson is identified in the $\jp$ recoil
system, {\em e.g.} in events of the type
$\ee\rt\jp D X$.  Figure~\ref{fig:Mrecoil_psi} shows
the distribution of masses recoiling against the $\jp$
for $\ee\rt\jp D X$ events where $M_X = m_{D^*}$ (top)
and $M_X = m_{D}$ (bottom).  There is an evident
$9.9\pm 3.3$~event signal for the $3940$~MeV state in the $D\bar{D^*}$
mass spectrum, with a statistical significance of $4.5\sigma$.
The signal level is the $D\bar{D}$ mass spectrum is
$4.1\pm 2.2$~events with a significance of only $2.1\sigma$.

This peak cannot be identified
with any known charmonium state.
An obvious guess is that it is either the $\chi_{c0}'$
or the $\etac''$.  However, $\chi_{c0}'\rt D\bar{D^*}$
is forbidden and, thus, ruled out.  Likewise
$\etac''\rt D\bar{D}$ decays are also forbidden, but,
since the $D\bar{D}$ ``signal'' is ambiguous, we can't
use this to rule out this assignment.  On the other hand,
an $\etac''$ assignment to the observed peak would
imply a $m_{\psi(3S)} - m_{\etac(3s)}$ mass splitting
of $\sim100$~MeV, about twice as large as the 
measured splitting for the $2S$ states.  This seems
unlikely.

The mass of this fourth peak is very similar to that 
of the $\omega\jp$ peak seen in $B\rt K\omega\jp$ and
described above, and a search for it
in the $\omega\jp$ decay channel is in progress.  In addition, we
are examining $B\rt K D\bar{D^*}$
decays for a $D\bar{D^*}$ component of 
the $\omega\jp$ enhancement.

\section{Summary}

We observe peaks near $3940$~MeV in the $\omega\jp$ mass distribution
from $B\rt K\omega\jp$ decays and in the recoil mass spectrum
in the inclusive annihilation process $\ee\rt\jp X$.  The latter 
peak is also seen in the exclusive process $\ee\rt\jp D\bar{D^*}$
and, thus, cannot be assigned to the $\chi_{c0}$ charmonium state.
At this stage, we cannot tell whether or not the state seen in $B$ decays 
and the one seen in inclusive $\jp$ production are one and the same.  
Further investigation is in progress.

We observe a $\sim 4\sigma$ signal for $X(3872)\rt\pipi\piz\jp$.
This is the first measurement of an $X$ decay mode other than $\pipi\jp$.
The $\pipi\piz$ invariant masses are strongly clustered above 750~MeV,
near the upper kinematic boundary; this is suggestive of a sub-threshold decay
via a virtual $\omega\jp$ intermediate state.  Such a decay, at near
the measured branching fraction, was 
predicted by Swanson based on a model where the $X(3872)$ is 
considered to be primarily a $D^0\bar{D^{0*}}$ hadronic 
resonance~\cite{swanson_1}.

The presence of the $X(3872)\rt\omega\jp$ decay process would establish the
Charge-Conjugation quantum number of the $X(3872)$ as $C=+1$.  This, in 
turn, would mean that
the $\pipi$ system in $X\rt\pipi\jp$ decay comes from the decay of a $\rho$
meson.  The large isospin violation implied by the near equality of
the $\rho\jp$ and $\omega\jp$ decay widths is difficult to accomodate
in a $c\bar{c}$ charmonium interpretation of the $X$, but a natural
consequence of the meson-meson bound state model point of view.

\section{Acknowledgements}
I thank the organizers of the APS Topical Group on Hadron
Physics for inviting me to give this talk.  My Belle colleagues and I
thank the KEKB group for the excellent operation of the
accelerator, the KEK cryogenics group for the efficient
operation of the solenoid, and the KEK computer group and
the NII for valuable computing and Super-SINET network
support.  We acknowledge support from MEXT and JSPS (Japan);
ARC and DEST (Australia); NSFC (contract No.~10175071,
China); DST (India); the BK21 program of MOEHRD and the CHEP
SRC program of KOSEF (Korea); KBN (contract No.~2P03B 01324,
Poland); MIST (Russia); MESS (Slovenia); Swiss NSF; NSC and MOE
(Taiwan); and DOE (USA).


\begin{thebibliography}{99}
\bibitem{conj} The inclusion of charge-conjugate modes
is always implied.
\bibitem{X_mass} This is the weighted average of all of the reported $X(3872)$
mass measurements.
\end{thebibliography}
\end{document}